# Paving Spin-Wave Fibers in Magnonic Nanocircuits Using Spin-Orbit Torque

Xiangjun Xing,[1,2] Philip W. T. Pong,[2] J. Åkerman,[3,4] and Yan Zhou[5,*]

[1]*School of Physics and Optoelectronic Engineering, Guangdong University of Technology, Guangzhou 510006, China*

[2]*Department of Electrical and Electronic Engineering, The University of Hong Kong, Hong Kong, China*

[3]*Department of Physics, University of Gothenburg, Fysikgränd 3, 412 96 Gothenburg, Sweden*

[4]*Material and Nano Physics, School of ICT, KTH Royal Institute of Technology, 164 40 Kista, Sweden*

[5]*School of Science and Engineering, Chinese University of Hong Kong, Shenzhen, 518172, China*

## Abstract

Recent studies have revealed that domain walls in magnetic nanostructures can serve as compact, energy-efficient spin-wave waveguides for building magnonic devices that are considered promising candidates for overcoming the challenges and bottlenecks of today's CMOS technologies. However, imprinting long strip-domain walls into magnetic nanowires remains a challenge, especially in curved geometries. Here, through micromagnetic simulations, we present a method for writing strip-domain walls into curved magnetic nanowires using spin-orbit torque. We employ Y-shaped magnetic nanostructures as well as an S-shaped magnetic nanowire to demonstrate the injection process. In addition, we verify that the Y-shaped nanostructures that incorporate strip-domain walls can function as superior spin-wave multiplexers, and that spin-wave propagation along each conduit can be controllably manipulated. This spin-wave multiplexer based on strip-domain walls is expected to become a key signal-processing component in magnon spintronics.

**Keywords:** magnonics, spin-wave waveguide, spin-wave multiplexer, spin-orbit torque, domain wall

---

[*]E-mail: yanzhou@hku.hk

**Introduction**

Spin-wave-based magnonic technologies [1–4] have great potential to overcome the obstacles met in electron-based CMOS technologies [5,6]; magnonic devices can provide significant performance benefits, such as enhanced throughput of information and reduced power consumption [3,7]. Information processing with wave properties can inherently avoid the extra heat production associated with electron transport found in conventional electronic circuits. Additionally, magnonic circuits are superior to CMOS circuits in implementing non-Boolean computations [8,9] for special types of data processing, such as pattern recognition [10], since parallel processing of signals can be readily realized via spin-wave interference [7].

Controllable, energy-efficient spin-wave propagation in magnonic circuits is a crucial step toward realizing practical spin-wave devices. In experimentally demonstrated prototype spin-wave devices [11–14], the Damon–Eshbach propagation geometry has been widely used because of the high group velocity and high efficiency of excitation (by the microstrip antenna) of the Damon–Eshbach spin waves [3,15]. Highly tunable spin waves and spin-wave beams can be excited using spin-transfer torque [16–18]. Recently, Duerr *et al.* presented a procedure for forming a self-cladding magnonic waveguide with narrow internal channels of as little as tens of nanometers in width, which can indeed enhance spin-wave transmission but cannot be applied to curved magnetic wires because of the nonlocal character of the field used to maintain the channel [19,20]. Later, Vogt *et al.* demonstrated continuous spin-wave propagation along either conduit of a magnonic multiplexer by applying an electric current along a metallic layer buried beneath the ferromagnetic layer: the Oersted field of the current aligns the magnetic moments orthogonally to each conduit [11,21]. Both approaches require an applied force (either an electric current [11,21] or a magnetic field [19,20]) to maintain the wave-guiding channels, with a considerable associated energy cost.

Alternatively, Garcia-Sanchez *et al.* have proposed using domain walls along a magnetic wire to guide spin waves and have numerically demonstrated spin-wave channeling along a curved wire [22]. Further, Xing *et al.* presented a method for reliably writing a domain wall into a magnetic wire, based on the spin-transfer torque of a current flowing through a delicately designed path [23]. Most recently, Wagner *et al.* experimentally verified that it is feasible to employ magnetic domain walls as spin-wave channels [24]. Under this scheme, the decay of spin-waves due to boundary scattering and intermodal scattering (caused by multimode confluence) can be reduced and, moreover, owing to the nonvolatility of domain-wall configurations, no additional energy needs be consumed to maintain the channels after formation. Thus, this waveguide scheme, which exhibits a prominent spin-wave channeling effect, appears very promising for novel magnonic devices [25]. However, injecting a long strip-domain wall (SDW) into a magnetic wire with curved geometry remains elusive. The domain-wall channels in the curved wires in Ref. [22] were preset via numerical conjectures; the domain-wall injection procedure proposed in Ref. [23] and that adopted in Ref. [24] seem valid only for straight magnetic wires. To allow broader applicability of the waveguide scheme in real magnonic devices, a procedure for writing long SDWs into magnetic wires with curvature is urgently needed.

In this work, we demonstrate how to form such a long SDW in curved magnetic nanostructures (Figs. 1–3), by using the driving effect of the emergent spin-orbit torque (SOT) on magnetic textures [26–30]. We stress that SOT resulting from spin-Hall effect (SHE) possesses the correct symmetry (Both the magnetization and electron spin orientation rotate locally with position along the length of a curved wire, but the resulting torque is always orthogonal to the length) to push the SDW head (the circular part of a SDW) when stabilizing the SDW ridge (the linear part of a SDW) (Fig. 4), and that a stable SDW cannot be realized using the scheme experimentally established in Ref. 29, because of intrinsic physical restrictions. Compared to the approach developed in Ref. [23], the present method can be applied to magnetic wires with various geometries (no longer limited to straight geometry) and to magnetic wires made of a magnetic insulator (no longer restricted to conductive magnetic materials). The procedure has been used to generate SDW-based spin-wave channels in Y-shaped nanostructures [11] (that can function as magnonic multiplexers, a frequently used signal-processing component) and also in an S-shaped magnetic wire, to show its robustness. This work should boost the popularity of magnonic devices that use domain walls as transmission channels for spin waves.

**Materials and Methods**

The basic structure of the domain-wall injection is shown in Figure 1. The main body is a Y-shaped nanostructure (middle panel, Fig. 1) made of the ultrathin multilayered films HM1/FM/AO$_x$/HM2 (bottom panel, Fig. 1), where FM is a ferromagnetic layer with perpendicular magnetic anisotropy (PMA), HM represents heavy-metal layers with strong spin-orbit coupling, and AO$_x$ is an oxide layer that insulates FM from HM2 [11]. Experimentally, the oxide layer can be formed by oxidizing the HM2 material [29,31] or other metallic materials [32–34] deposited on FM, or by directly depositing a dielectric oxide [27,30] on FM. The asymmetric interfaces of the FM layer combined with the strong spin-orbit coupling in HM1 can induce a large interfacial Dzyaloshinskii–Moriya interaction (DMI) in the ferromagnet [35], which is required to stabilize SDWs in curved geometries, as will be shown later. Narrow pads (A, B, C, and D) for domain-wall nucleation at each end of the FM layer (top panel, Fig. 1) are used to form seed reverse domains. An in-plane charge current flowing in HM2 along S1–S3 or S2–S3 generates a pure transverse spin current in the perpendicular direction via the spin-Hall effect [36,37], which in turn exerts spin-orbit torques on the magnetization in FM [38]. We intentionally insert a thin $AO_x$ layer between FM and HM2 to prevent charge current from flowing into the ferromagnet, thus eliminating the Zhang-Li spin-transfer torques (STT) [39,40] in the ferromagnet (Actually, inserting the insulating layer is not essential, because the SDW injection is not significantly influenced by the Zhang-Li torques accompanying the SHT; see Supplemental Material [41]). Depending on the current path used (S1–S3 or S2–S3), a SDW can be written into the top or bottom arm.

We performed micromagnetic simulations to study SDW injection and spin-wave propagation along SDWs in Y-shaped nanostructures by numerically solving the motion equation of the magnetization—that is, the modified Landau-Lifshitz-Gilbert equation [42,43] with spin-orbit torques [38,44],

$$\partial \mathbf{m}/\partial t = -\gamma (\mathbf{m} \times \mathbf{H}_{eff}) + \alpha (\mathbf{m} \times \partial \mathbf{m}/\partial t) + \mathbf{T}_d + \mathbf{T}_f,$$

where $\mathbf{m} = \mathbf{M}/M_s$ is the unit vector along the magnetization $\mathbf{M}$ and $M_s$ is the saturation magnetization; $\mathbf{H}_{eff} = -(1/\mu_0)\delta E/\delta \mathbf{M}$ is the effective field in the ferromagnet with $\mu_0$ denoting the vacuum permeability and $E = E_d + E_u + E_x + E_{DM} + E_Z$ the total energy density, including the magnetostatic, anisotropy, exchange, DMI, and Zeeman energy contributions; $\mathbf{T}_d$ and $\mathbf{T}_f$ represent the damping-like [45] and field-like [46] torques, respectively.

The LLG micromagnetic simulator (a commercial GPU code) [47] was used to implement all the simulations, in which only the FM and HM2 layers were explicitly incorporated, as in Ref. [43]. We did not directly include the HM1 and $AO_x$ layers in the simulations, but instead simply took account of their physical effects—that is, the DMI caused by HM1 and the insulation of FM from HM2 enabled by $AO_x$. Both the FM and HM2 layers are $d_{FM} = d_{HM2} = 1$ nm in thickness. The width of the wire for all samples is $w = 100$ nm, and the wire length varies with sample geometry. We examined the SDW injection process and its stability over a wide range of values of $M_s$, $K_u$, and $D$, to account for the sensitivity of the material parameters to interface properties and layer thickness [34,48,49]. The results are based on the following material parameters, unless otherwise specified: $M_s = 580$ kAm$^{-1}$, exchange stiffness $A = 15$ pJm$^{-1}$, perpendicular magnetocrystalline anisotropy $K_u = 0.8$ MJm$^{-3}$ (Thus, the effective uniaxial anisotropy is $K_{eff} = 0.6$ MJm$^{-3}$, as determined from $K_{eff} = K_u - (1/2)\mu_0 M_s^2$), DMI strength $D = 2.0$ mJm$^{-2}$, and Gilbert damping constant $\alpha = 0.02$. These parameters used correspond to the experimental values reported for Pt/Co/AlO$_x$ [50], Pt/CoFeB/MgO [30], and Ta/CoFeB/TaO$_x$ [31,51] systems. For computation, each sample was divided into regular meshes of 2×2×1 nm$^3$ in size, which is much smaller than the exchange length $l_{ex} = \sqrt{2A/\mu_0 M_s^2} \approx 8.4$ nm (the maximum length within which the magnetization can be kept uniform by the short-range exchange interaction), and open boundary conditions were used. For simulations with spin-orbit torques, the spin-Hall angle was assumed to be $\Phi_H = 0.13$ [52], the spin polarization of the carriers in the FM layer $P = 0.4$, and a series of values of the Rashba parameter ($\alpha_R$) were considered [53,54] to examine the influence of the Rashba torque (RsbT; We consider that SHE only contributes to the damping-like torque while the Rashba effect only induces the field-like torque, as implemented in Ref. [43]. The largest Rashba parameter examined gives the Rashba torque 1.5 times as strong as the spin-Hall torque (SHT) for $J_{FM} = J_{HM2}$). To check the contribution of the Zhang-Li torques in the SDW injection driven by SHE, equal current densities in FM and HM2, *i.e.*, $J_{FM} = J_{HM2}$, and $\beta=\alpha$ ($\beta$ is the nonadiabaticity factor [39,40]) were implemented in corresponding simulations. The effect of pinning centers on the SDW injection was examined by considering unrealistically high concentrations (1% and 5%) of impurities inside the FM layer, which are modeled as randomly distributed sites over the simulation space with $D_{(Impurity)} = -D_{(Non-impurity)}$ [30,55]. We also included a Langevin random field in the effective field to test the thermal-fluctuation effects on the SDW injection.

An electric current, $I_{dc}$, was fed to S1–S3 or S2–S3 yielding a quasiuniform current density in the wire away from the transition section (Note that the electric current is confined to HM2 when the oxide layer is present; thus $J_{HM2} = I_{dc}/wd_{HM2}$). The real current distribution along S1–S3 in the Y-shaped nanostructure for $I_{dc}$ = 300 μA is overlain onto the top view in Figure 1 as a contour plot coded into $J_x$ (the *x*-component of the current density), indicating that the current distribution is slightly inhomogeneous around the transition region. To mimic experimental conditions, such real current distributions [55] were used in all of our simulations, instead of uniform ones [56,57], and the Oersted fields given by these current distributions were also incorporated in all relevant simulations.

**Results**

Writing SDWs into magnetic nanostructures is the main focus of the present research. Figure 2 shows the steps for imprinting an SDW. First, a quasiuniform single domain (Fig. 2a) is generated, which can be realized numerically by relaxing an artificial spin configuration with $m_z$ = -1 to static equilibrium. Experimentally, a sufficiently strong magnetic field along -*z* can be used to set up such a spin configuration. Second, a seed domain wall is injected into the narrow pad A. To do this numerically, we apply a sinusoidal field pulse [56], $\mathbf{H}_i = \mathbf{H}_0 \sin(2\pi f_i t)$, with H$_0$ = 1.4142 T ($\mathbf{H}_0$ is in the plane normal to the pad length and 45° away from the *xy*-plane), $f_i$ = 2.5 GHz, and $t$ = 200 ps to the outmost 50×$w_N$ nm$^2$ of pad A ($w_N$ is the pad width). Figure 2b shows the spin configuration at a time when $\mathbf{H}_i$ runs for 200 ps. It is worth noting that, experimentally, a seed reverse domain can be injected by using a local spin valve formed on the pad [58]. Third, a direct current ($I_{dc}$ in Fig. 1) is applied to the top conduit from S1 to S3, and as a result a transverse vertical spin current is injected into the FM layer. This spin current imposes a spin torque on the seed domain wall via SHE [37] (The Rashba effect may occur here but will not significantly affect the SDW injection, if the relative strength of the RsbT to SHT is below 0.5; see Supplemental Fig. S1 [41]).

The SHT drives the seed domain wall to move forward and enter the arm (Fig. 2c). After entry, the seed domain wall, initially aligning along the transverse direction (Fig. 2b), evolves into an SDW with its ridge aligning along the arm (Fig. 2d). It is intriguing that SHT only moves the head of the SDW, but does not shift the ridge of the SDW. We will show that this is because the magnetization in the SDW ridge (at the SDW head) is parallel (perpendicular) to the electrons' spin orientation defined by SHE, and therefore feels a vanishing (considerable) torque. At ~1975 ps, the SDW head passes through the bent section of the top conduit and enters into the base. At ~3949 ps, the SDW head reaches the left end and an intact SDW is written into the top conduit. Next, by withdrawing the electric current and relaxing the whole system to static equilibrium, a stable SDW is obtained, as shown in Figure 2g. Similarly, an SDW can be imprinted on the bottom conduit from pad B (Fig. 2h).

In Figure 2, the initial domain is downward, the seed reverse domain is upward, and an electric current along S1–S3 successfully writes an SDW into the top conduit. We now begin the process from an upward initial domain (Fig. 3a): clearly, we need to create a seed domain with downward magnetization (Fig. 3b) using a field pulse. Once a seed domain (and then a seed domain wall) forms, an electric current along S1–S3 is turned on. Figure 3c–f illustrates the dynamic evolution of the seed domain wall after the application of current; this is very chaotic and hence is totally different from the process in Figure 2c–f. In this case, the seed domain wall can enter into the arm (Fig. 3c), as in Figure 2c, but it cannot sustain a profile as an SDW (compare Figs. 2d,3d). In fact, the early domain wall in Figure 3c already extends transversely and soon contacts the other edge of the arm (Fig. 3e). Thereafter, disordered domain patterns (Fig. 3f) appear and evolve dynamically in the arm (Supplemental Movie S1b [41]) until the current is turned off. Figure 3c–f indicates that it is impossible to inject an SDW into the Y-shaped nanostructure from a downward seed domain situated in pad A.

In Figures 2 and 3, pad A is used to inject an SDW. SDW injection can in fact be initiated from any one of pads A, B, C, and D. However, as in the case using pad A, only the seed domain with a specific magnetization orientation can result in the successful imprinting of an SDW, and the seed domain with an opposite orientation will cause chaotic dynamics (see Supplemental Movies S1a–h [41] for details). The dependence of SDW injection on the magnetization orientation of a seed domain can be understood as follows: assuming that an SDW can also be injected into the arm from pad A with the wrong magnetization orientation, the magnetization orientation in the ridge of the SDW would be antiparallel to the spin orientation of the electrons polarized through SHE (see Fig. 3g). However, only if the magnetization in the ridge of an SDW is parallel to the spin orientation of the polarized electrons (see Fig. 3h) can the strip-like profile of the SDW be maintained and the SDW continues to exist in the arm (Fig. 4d); otherwise, the strip-like profile cannot be dynamically stabilized, and disordered domain patterns form instead (Fig. 4e). This picture is systematically corroborated in Figure 4.

Figure 4a shows a strip domain (pinned at the left end of the wire) that is enclosed by two parallel SDWs and a semicircle domain wall (forming a half-skyrmion also called meron [56,59]).

The pinned strip domain with paired SDWs will split into two strip domains (Fig. 4b), each of which has an isolated SDW parallel to the edge and a SDW head (as a quarter-skyrmion) touching the edge, if the wire is cut along its middle line. With stripe domains as the seeds, current-driven generation of skyrmion bubbles was elegantly demonstrated in the recent literature [29], and the spatially divergent current across the paired SDWs induced by the constriction was reckoned to be responsible for the skyrmion generation. However, according to the latest theory, specially developed to explain the experimental results in Ref. [29], Lin [60] revealed that a strip domain with paired SDWs cannot sustain and exist in the system, once a current is applied, because the half-skyrmions (acting as the SDW head) at the ends of a stripe domain will move transversely due to the Magnus force associated with the finite topological charge of a half-skyrmion. Consequently, a stripe domain subjected to a current will be distorted in a zigzag manner and eventually break into massive skyrmions.

Therefore, inherently, the strip domain with paired chiral SDWs (as shown in Fig. 4a) cannot be written into a magnetic wire by current-induced SHT because of the physical limit identified by Lin [60]. Below, we show that there exists the other intrinsic mechanism that makes it impossible to inject such a strip domain with paired chiral SDWs into a wire by SHT (Supplemental Fig. S5 [41]). The reason is that one of the two chiral SDWs has the magnetization antiparallel to the electrons' spin orientation defined by SHE (as shown in Fig. 4a), and will be destabilized by the resulting SHT immediately after the current application. Figure 4c illustrates the current-driven dynamics of an open strip domain without including the half-skyrmion head (*i.e.*, the part encircled by the box in Fig. 4a). It is seen that, one of the paired SDWs (here, the bottom SDW) begins to deform once the current is applied, and is heavily distorted and intersects the bottom edge at 148 ps after the current application. Its subsequent chaotic motion will in turn destroy the other SDW (here, the top SDW) and result in disordered domain patterns in the nanowire. When the two SDWs are spatially separated (Fig. 4d,e), no interplay happens to them again. As a consequence, under the current action, one SDW sustains its initial geometry and always resides in the wire (Fig. 4d), while the other SDW distorts and finally breaks into pieces (Fig. 4e). The results in Figure 4c-e serve as rigorous proofs for the hypothesis stated above and illustrated in Figure 3g,h, and provide additional insight, along with Lin's latest theory [60], into the current-driven dynamics of strip domains.

As shown in Figure 4b, after division, each SDW consists of the SDW ridge (linear part) and the SDW head (circular part). The SDW head, which can be approximately regarded as a quarter-skyrmion, has nonzero topological charge. Thus, under the current as displayed in Figure 4b, the bottom SDW cannot maintain its profile and will fluctuate irregularly, as what the SDW behaves in Figure 3f. However, under the same current, the top SDW can keep its general shape and gradually grow along the edge. This progressive elongation of the SDW benefits from the stabilization of the SDW ridge and the steady longitudinal motion of the SDW head along the edge. In fact, there is a transverse Magnus force acting on the quarter-skyrmion-like SDW head as a result of its finite topological charge [59] (see Supplemental Fig. S6 [41]). Nevertheless, here, the SDW head is tied to the edge, so that the confining force coming from the edge cancels the Magnus force, and the transverse motion of the SDW head is suppressed completely, avoiding the zigzag distortion of the SDW and the creation of associated topological charge as described in Ref. [60]. That is to say, by splitting the strip domain, with paired chiral SDWs, into separate SDWs, each attached to an edge, the two independent physical mechanisms, responsible for the destabilization of a strip domain, can be simultaneously deactivated, and moreover, a single SDW can be imprinted into a wire by using SHT.

In Ref. [29], strip domains were indeed observed to exist in the studied micrometer-sized samples subjected to ultralow current densities. However, the observed strip domains, which were deemed to be pinned and thereby extrinsically stabilized by the randomly distributed impurities inside the samples [60], exhibit an irregular zigzag profile, making the SDWs not suitable for guiding spin waves as smooth fibers. Furthermore, stabilization of the zigzag SDWs by impurities will become invalid for fast nanoscale devices that demand high-speed operation under reasonably high current densities.

We examined the applicability of the injection procedure with respect to geometric variation in the sample, including the opening angle between the base and the arms, the corner shape around the transition region, and the pad width of the Y-shaped nanostructure, and found that the injection procedure is generally valid throughout the considered geometries and also for the sample with vanishing DMI. The details are presented in the Supplemental Movies S1–S6 [41].

The equilibrium domain patterns in the 90° Y-shaped nanostructure, relaxed from the as-written SDW for various values of $D$ (Fig. 5), clearly indicate that the DMI strength must be in the proper range to stabilize the SDW at static equilibrium (Fig. 5c–e). For subthreshold $D$ values, the as-written SDW will transform into a multidomain texture (Fig. 5a) or will disappear, resulting in a single domain (Fig. 5b); for suprathreshold $D$ values, it will break into the labyrinthine worm-like texture shown in Figure 5f. The stability of SDWs in a Y-shaped nanostructure with respect to DMI strength is different from the situation for a straight magnetic wire, where SDWs are stable even for $D = 0$, as reported in Refs. [23,24]. This implies that the stabilization of an SDW in a curved wire requires a sufficiently strong DMI. In the present study, the Y-shaped nanostructure includes a base, two arms, and a transition section, and the transition region between the base and the arm is a segment of a magnetic ring. Therefore, the dependence of SDW stability in a Y-shaped nanostructure on $D$ should be due to the presence of the transition region. For our Y-shaped nanostructures, the transition region is a 100-nm-wide arc with an outer radius of 400 nm. We thus examined the stability of an SDW in such a 1/4-arc against $M_s$, $K_u$, and $D$. The results displayed in Supplemental Figures S7–S11 [41] indicate that an SDW may be stabilized in the curved wire (1/4-arc) at static equilibrium over a broad range of values of $M_s$, $K_u$, and $D$.

We now turn to the usefulness of the Y-shaped nanostructures, with controllably written SDWs, as spin-wave multiplexers. First, we examined spin-wave propagation in a 60° Y-shaped nanostructure with an SDW placed in the top or bottom conduit, or without an SDW. Figure 6 displays the propagation patterns of spin waves at 30 GHz. It can be seen that the spin waves are guided along an SDW in the top or the bottom conduit (Fig. 6a,b), but do not exist inside the nanostructure without an SDW (Fig. 6c). Here, regarding the channeling effect, the bottom arm in Figure 6a and the top arm in Figure 6b are equivalent to the corresponding arms in Figure 6c. Figure 6d,e plots the spin-wave amplitudes in symmetric zones of the 60° nanostructure, with an SDW included in either arm, over a frequency range of 80 GHz, and suggests that the top and bottom SDWs are analogous as far as the channeling effect is concerned. The spin-wave strength in the arm with an SDW is more pronounced (approximately 1–2 times stronger) than that in the arm without an SDW. Note that a data point on the curves in Figure 6d,e represents the spin-wave amplitude averaged over the area of the dashed box (Fig. 6a,b); the actual difference between the spin-wave amplitude at the SDW position and that at the corresponding position without an SDW is thus underestimated. These results indicate that the Y-shaped nanostructures can behave as current-controlled spin-wave multiplexers, operating over a broad frequency band, which are expected to be of high energy efficiency owing to the nonvolatility of the rewritable spin-wave guiding channels. Such low power consumption, as a key figure of merit for device applications [61], is difficult to achieve in the multiplexer based on Damon-Eshbach spin-wave channels maintained by current-induced Oersted field, recently proposed in Ref. [11].

In a further step, spin-wave propagation along SDWs inside the Y-shaped nanostructures with different opening angles was examined; the results are shown in Figure 7. Clearly, for all three opening angles, the SDWs in the Y-shaped nanostructures can channel spin waves, indicating that the SDW-based spin-wave multiplexer can operate over a wide angle range. Although for all the three opening angles the spin waves can travel smoothly along the SDWs, the spin waves for larger opening angles decay faster after passing through the corner, consistent with Ref. [11]. That is to say, the larger the opening angle is, the heavier the spin-wave attenuation with the propagation distance; this is possibly because the spin waves in multiplexers with larger opening angles experience stronger intermodal scattering or boundary scattering [21,62–66].

To excite spin waves, we adopted the inductive method based on a strip-line antenna, which produces a magnetic field with a symmetric profile relative to the SDW elongation axis. With this inductive scheme, only spin waves with $2n$ ($n = 1, 2, 3…$) nodes can be activated, and the excitation efficiency of a mode is proportional to $1/(2n-1)^2$, where $2n-1$ is the order of the excited mode [66,67]. Consequently, the ratio between the nominal excitation strengths of the third-order and first-order (fundamental) modes is 1/9. In particular, only the fundamental mode is strongly excited at the antenna. When the fundamental mode runs into the bent section of an SDW, the $2n$-order mode can be activated because of the translation-symmetry breaking [63,66]. In this way, the fundamental mode can scatter into higher-order modes [68]. An increased opening angle makes a transmission channel undergo stronger bending, which in turn leads to enhanced intermodal scattering and thus enhanced attenuation of the fundamental mode.

On the other hand, the distance between the SDW and the top edge decreases as the opening angle increases, as seen in Figure 7. This reduced separation has two effects: first, the fiber mode (the mode confined in the SDW) [23] will mix with the edge mode [62,69,70] because of the overlap of their spatial profiles; second, the fiber mode will be scattered by the boundary defects [62], such as edge roughness (In our simulations, the staircase along the edge resulting from the finite-difference meshing can be regarded as weak edge roughness). The modal scattering between the fiber and edge modes, as well as boundary scattering, become two additional energy-dissipation channels for the fundamental fiber mode.

To ensure that the proposed SDW injection procedure is generally valid for curved wires, we used it to write an SDW into an S-shaped wire, whose curved parts are 100-nm-wide arcs with an outer radius of 400 nm. The whole process, resembling that in Figure 2, is illustrated in Figure 8. The initial state is a single domain with $m_z = +1$. A seed reverse domain is formed in a nucleation pad by using a magnetic field pulse, and then an electric current is feed into the HM2 layer of the wire from the lower-left to the upper-right terminal. The zero picosecond point in Figure 8b marks the time when the electric current is switched on. Once the current is applied, the seed domain wall moves into the wire and becomes an SDW (Fig. 8c). Under the continuous action of the current, the SDW moves forward and ultimately extends over the entire length of the wire (Fig. 8d–f). A variation in strip-domain width with position (Fig. 8g) also occurs to the SDWs in the Y-shaped nanostructures (Fig. 2f,g), which can be ascribed to compromise between the various competing energy terms.

**Discussion**

For the systems considered here, Néel domain walls are the preferred type for rendering the most efficient domain-wall motion by SHT, as found in Refs. [27,43]. During the SDW injection process (Fig. 2), the effective field ($\mathbf{H}_{SH} \propto \mathbf{m}\times\boldsymbol{\sigma}$ [43]) associated with SHT ($\mathbf{T}_d = -\gamma\tau_H(\mathbf{m}\times\boldsymbol{\sigma}\times\mathbf{m})$, where $\tau_H = \hbar J\Phi_H/2eM_s d_{FM}$, $\boldsymbol{\sigma} = -\hat{\mathbf{J}}\times\hat{z}$, $\hbar$ is the reduced Planck constant, $e$ the elementary charge, $d_{FM}$ the thickness of the FM layer, $\hat{\mathbf{J}}$ the unit vector in the current direction, and $\hat{z}$ the unit vector along the $z$ axis; see Refs. [37,38]) exerting on the magnetization at the SDW head is aligned with the magnetization of the newly formed reverse domain, and thus the SDW head moves forward through the continuous expansion of the reverse domain along an edge of the wire. However, the effective field in the SDW ridge vanishes because of the parallel alignment of the electron spins and the magnetization in the SDW ridge (Fig. 3g), so that after formation, the SDW ridge does not experience a torque and thus stops.

The Rashba effect has been suggested as the main driving force for domain-wall motion in a Pt/Co/AlO$_x$ wire [71]. It is thus essential to identify whether the Rashba effect plays a key role in the present case. But, as we have seen from simulations, the Rashba torque ($\mathbf{T}_f = -\gamma\tau_R(\mathbf{m}\times\boldsymbol{\sigma})$, where $\tau_R = JP\alpha_R/\mu_B M_s$ and $\mu_B$ is the Bohr magneton; see Ref. [44]) coexisting with SHT does not significantly change the SDW injection process as long as the relative strength of RsbT to SHT, $\tau_R/\tau_H$, is not higher than 0.5 (see Supplemental Material [41] for details). The Zhang-Li torque accompanying SHT makes even smaller contribution to the SDW injection than RsbT, and only alters the effective velocity of the SDW head (see Supplemental Material [41]). Notably, the proposed procedure can work at room temperature and is not sensitive to impurities inside the sample even for an ultrahigh impurity concentration of 5% (Supplemental Fig. S4 [41]).

Different from the approach presented in Ref. [23], which would seem only to apply to straight magnetic wires, the method proposed here can work well for SDW injection in long magnetic wires with curvature (even at room temperature, as shown in Supplemental Movie S7 [41]). On the other hand, the method used in Ref. [23] requires a spin valve or magnetic tunnel junction to produce a spin-polarized current, so the device structure is complex. By contrast, the present writing scheme does not require excessive units to generate a spin current. Another difference between the two injection schemes is that the former method is based on magnetization switching to form a reverse domain, whereas the present method relies on the motion of a domain-wall head to yield a reverse domain (Once the reverse domain forms in the background domain, the injection of an SDW, situated between the initial and reverse domains, is achieved). From the viewpoint of applications, curved components [21,63,64,66,72] will be an unavoidable building block in functional magnonic circuits [7,10,11]. The present method is well suited for writing SDWs into realistic circuits with curved parts to form fiber-type spin-wave waveguides [22–24].

We pass the electric current only to the HM2 layer, rather than to the entire thickness, by using $AO_x$ as an isolation element to avoid the conventional Zhang-Li spin torques and thus ensuring that only the SOT exists in the FM layer. By doing so, the computations can be greatly simplified, since the relative strength of the nonadiabatic and adiabatic torques is not definitely known (still under debate actually). Experimentally, in principle, eliminating the Zhang-Li torques is not required for SDW injection in the HM1/FM/$AO_x$/HM2 system, because they contribute negligibly to domain-wall displacement in ultrathin multilayer nanostructures, as argued in Ref. [27] and demonstrated in Supplemental Fig. S2 [41].

The drawback of the present SDW injection method is that it is invalid for all domain-wall types other than Néel walls (For example, it is not compatible with Bloch or transverse walls). It can thus only be used in material systems with PMA and DMI that favor the Néel-type domain wall in a magnetic wire [73]. The recently reported ultralow magnetic damping $\sim 10^{-4}$ (previously attained only for ferrimagnetic YIG films [74,75]) of sputter-deposited polycrystalline CoFe films with Cu/Ta seed and capping layers [76] can exhibit all the required PMA, DMI, and SOT [27], and therefore might bring an unprecedented opportunity for the demonstrated SDW injection procedure (Besides 0.02, we also tested other damping values, 0.01, 0.005, and 0.001, for which the injection processes do not distinguish from each other) and the established spin-wave guiding scheme based on SDWs. However, the commonly used material systems for present-day magnonic applications are low-damping Py [1,3] (metallic) and YIG [77] (insulating) films without PMA, for which the proposed injection scheme will fail because of mismatching between domain-wall configuration (transverse domain wall) and SOT [43].

Although the structure of our multiplexers is similar to that in Ref. [11], and we also use an electric current to set a spin-wave channel, our waveguide mechanism (the optic-fiber-like waveguide [22–24]) is different; in particular, no current is needed to maintain the spin-wave channels in our multiplexers after imprinting, leading to substantially reduced power consumption. Additionally, the energy benefit and its applicability to curved samples might make our waveguide outperform the one exploited in Ref. [19,20].

In conclusion, we have proposed a robust method for writing strip-domain walls into magnetic nanostructures patterned from an ultrathin multilayer film with perpendicular magnetic anisotropy, the interfacial Dzyaloshinskii–Moriya interaction, and the spin-Hall effect, and even containing high-concentration impurities. Apart from straight wires, the method can be used for curved samples even at room temperature. Moreover, the spin-wave waveguide and multiplexer based on strip-domain walls are energy efficient, compared to previously established ones. These findings may drive them to become prototypical spin-wave devices in magnonics. We also identify an emergent physical mechanism for the stabilization/destabilization of a domain wall under the spin-Hall torque, which might lead to novel operation concepts in domain-wall-based memory and logic devices.

**Acknowledgements:** Y.Z. acknowledges support by the National Natural Science Foundation of China (Project No. 1157040329) and Shenzhen Fundamental Research Fund under Grant No. JCYJ20160331164412545. X.J.X. acknowledges the support of the National Natural Science Foundation of China under Grant No. 11104206.

**Conflict of Interests:** The authors declare no competing financial interests.

**Figure captions**

Figure 1. **Device structure and control circuit.** The Y-shaped nanostructure is patterned from a multilayer film, HM1/FM/AO$_x$/HM2. Each end of the FM layer has two narrow pads 150 nm long and 20 nm wide for nucleating seed domain walls. A switch connects leads S1–S3 and S2–S3 to the direct-current source $I_{dc}$. The current is confined in the HM2 layer by the insulating AO$_x$ layer. The current profile between S1–S3 overlaid on the FM layer exhibits an inhomogeneity at the transition region because of the variation in the wire width. $J_x$ is the *x*-component of current density and the red arrows denote in-plane current directions. Here, the opening angle between the two arms, symmetrical relative to the horizontal base, is 90°.

Figure 2. **SDW injection process starting from a right seed domain.** (a) Initial single-domain state. (b) Seed reverse domain formed at nucleation pad A. Current is switched on at 0 ps when the seed domain is just formed. (c–f) Transient-state SDWs at indicated times after current application. (g) Static SDW after relaxation from 3949 ps. (h) Static SDW in the bottom conduit. The complete dynamic process is shown in Supplementary Movie S1a [41].

Figure 3. **Injection process initiated from a wrong seed domain and microscopic origin of SDW stabilization and destabilization.** (a) Initial single-domain state. (b) Seed domain formed at nucleation pad A. Current is switched on at 0 ps when the seed domain just forms. (c–f) Transient states at indicated times after current application. No SDW can be written into the arm because of chaotic dynamics. The complete dynamic process is shown in Supplementary Movie S1b [41]. (g) Spin orientation in the ridge of the fictitious SDW is opposite to electrons' spin orientation set by the SHE [37], and therefore the SDW ridge will be destabilized by SHT, leading to disordered domain patterns. (h) Spin orientation in the ridge of the right SDW is parallel to electrons' spin orientation defined by the SHE, and thus no SHT acts on the SDW ridge. The SDW head feels a SHT and moves forward.

Figure 4. **A physical mechanism responsible for domain-wall stabilization and destabilization.** (a) A strip domain, pinned at the left side of a wire, has two paired linear SDWs and a semicircle domain wall that forms a half-skyrmion with finite topological charge (denoted by '+'). **J** is the current density, **σ** stands for the electrons' spin orientation given by SHE, and **m** represents the magnetization direction in the domain wall. (b) The split strip domains, each with a single SDW: the linear SDW ridge is parallel to the edge, and the bent SDW head (quarter-skyrmion; surrounded by the box) is attached to the edge. Here, the SDW head still has nonzero topological charge. Dynamics of (c) the paired SDWs, (d) the isolated SDW where **m** // **σ**, and (e) the isolated SDW where **m** is antiparallel to **σ**, under current-induced SHT. Zero picosecond corresponds to the time when the current is applied to the equilibrium spin configuration.

Figure 5. **Static spin configuration in a Y-shaped nanostructure as a function of *D*.** Each pattern is obtained by relaxing an as-written SDW (as shown in Fig. 2f) in the top conduit. $A$=15 pJm$^{-1}$, $M_s$=580 kAm$^{-1}$, and $K_u$=0.8 MJm$^{-3}$. The opening angle of the nanostructure is 90°.

Figure 6. **Y-shaped nanostructure, with controllably written SDWs, as spin-wave multiplexer.** (a) Spin-wave transmission along the top conduit with an SDW. (b) Spin-wave transmission along the bottom conduit with an SDW. (c) Spin-wave transmission prohibited in the nanostructure without including SDWs. (d,e) Comparison of spin-wave amplitudes on the two arms with various spin configurations. Plots d and e correspond to plots a and b, respectively. Spin waves are excited at the antenna. The spin-wave frequency is 30 GHz. The opening angle of the Y-shaped nanostructure is 60°.

Figure 7. **Spin-wave transmission in a multiplexer dependent on opening angles.** The opening angles for the multiplexers in (a), (b), and (c) are 30°, 60°, and 90°, respectively. Spin waves are excited at the antenna. The spin-wave frequency is 30 GHz.

Figure 8. **SDW injection into an S-shaped nanowire.** (a) Initial single-domain state. (b) Seed domain formed at a nucleation pad. Current is switched on at 0 ps when the seed domain wall just forms. (c–f) Transient-state SDWs at indicated times after current application. (g) Static SDW after relaxation from 12089 ps.

FIG. 1

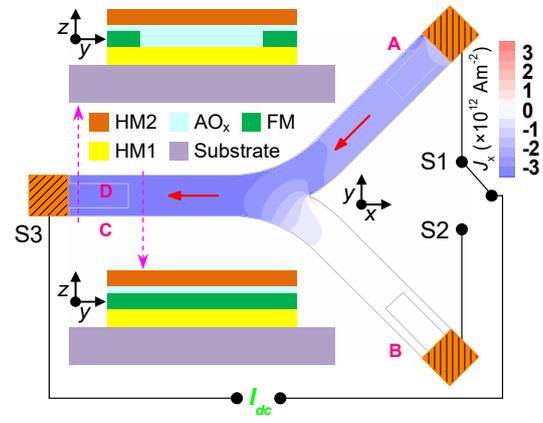

FIG. 2

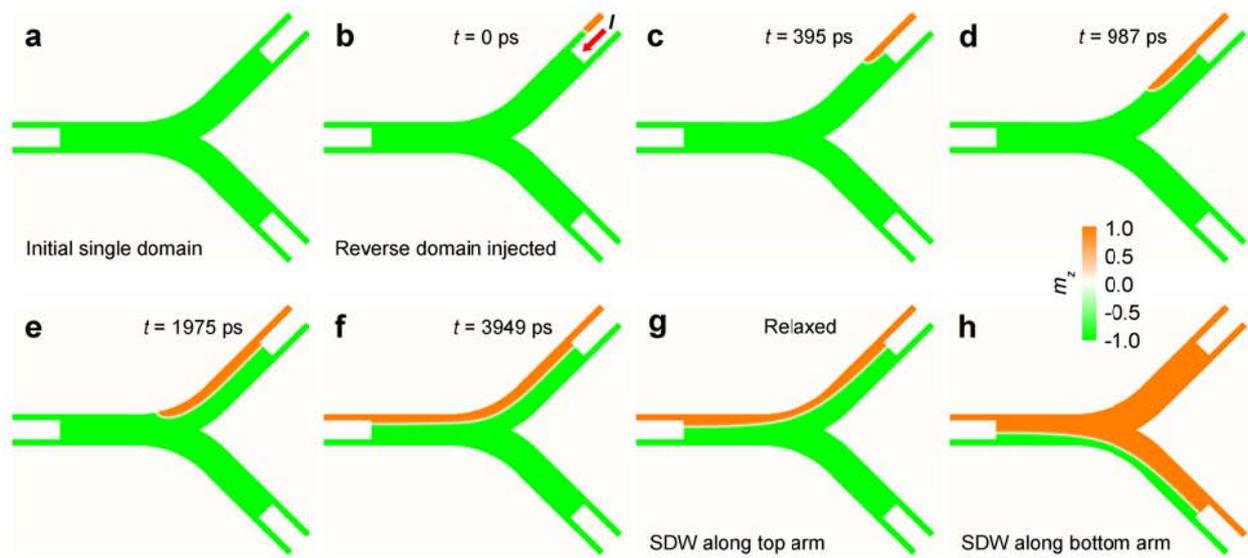

FIG. 3

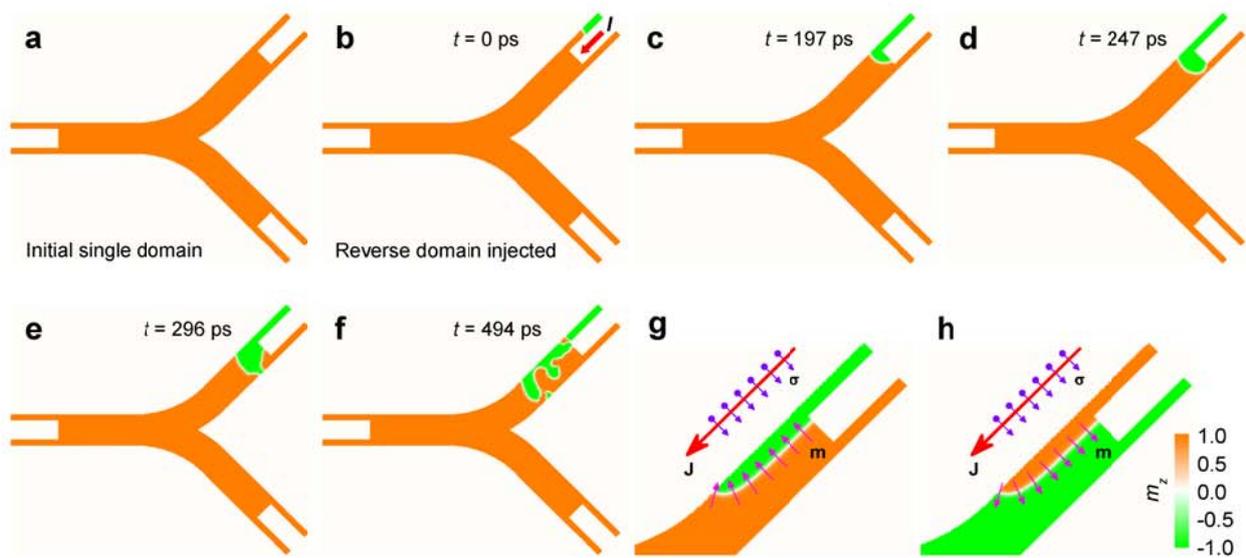

FIG. 4

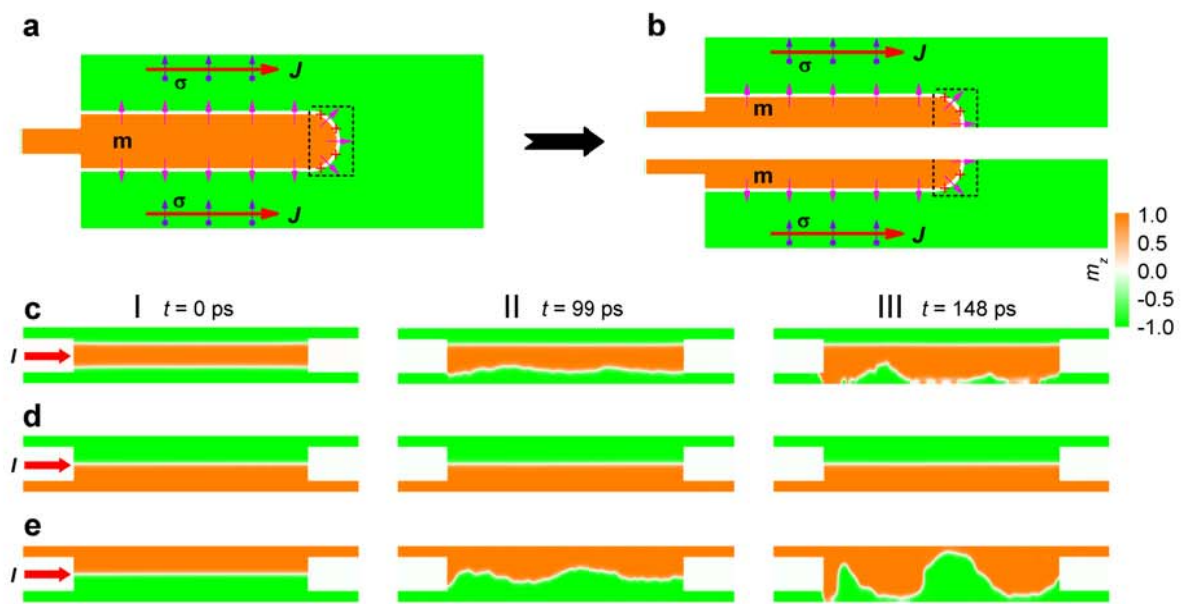

FIG. 5

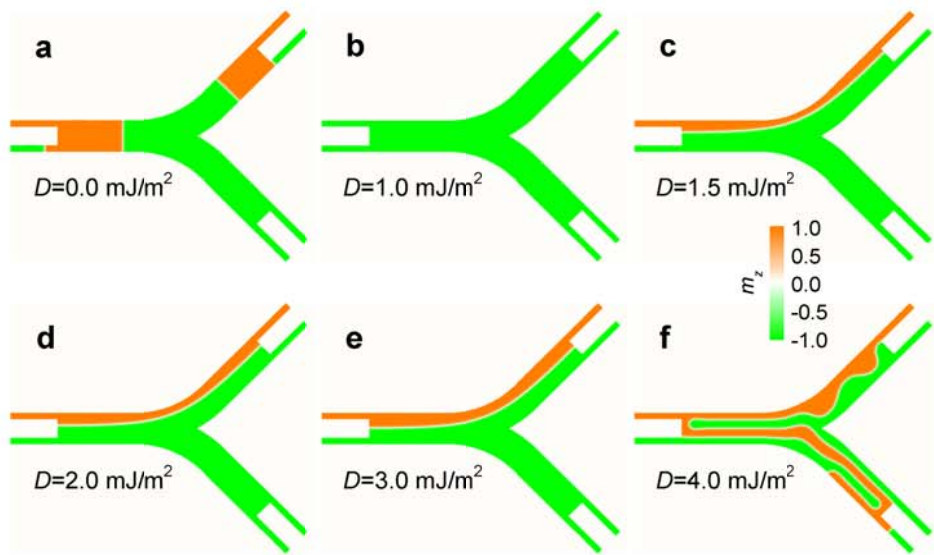

FIG. 6

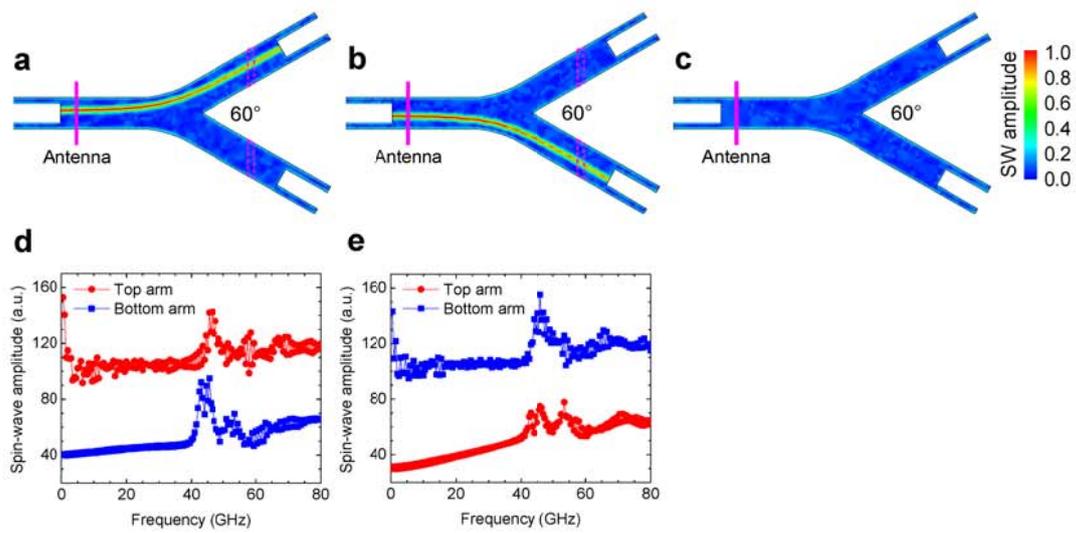

FIG. 7

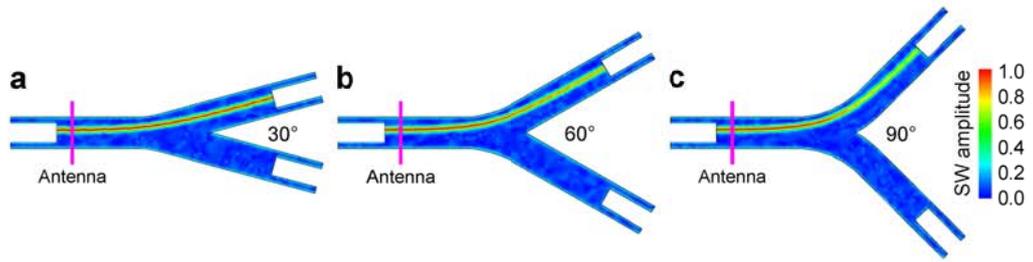

FIG. 8

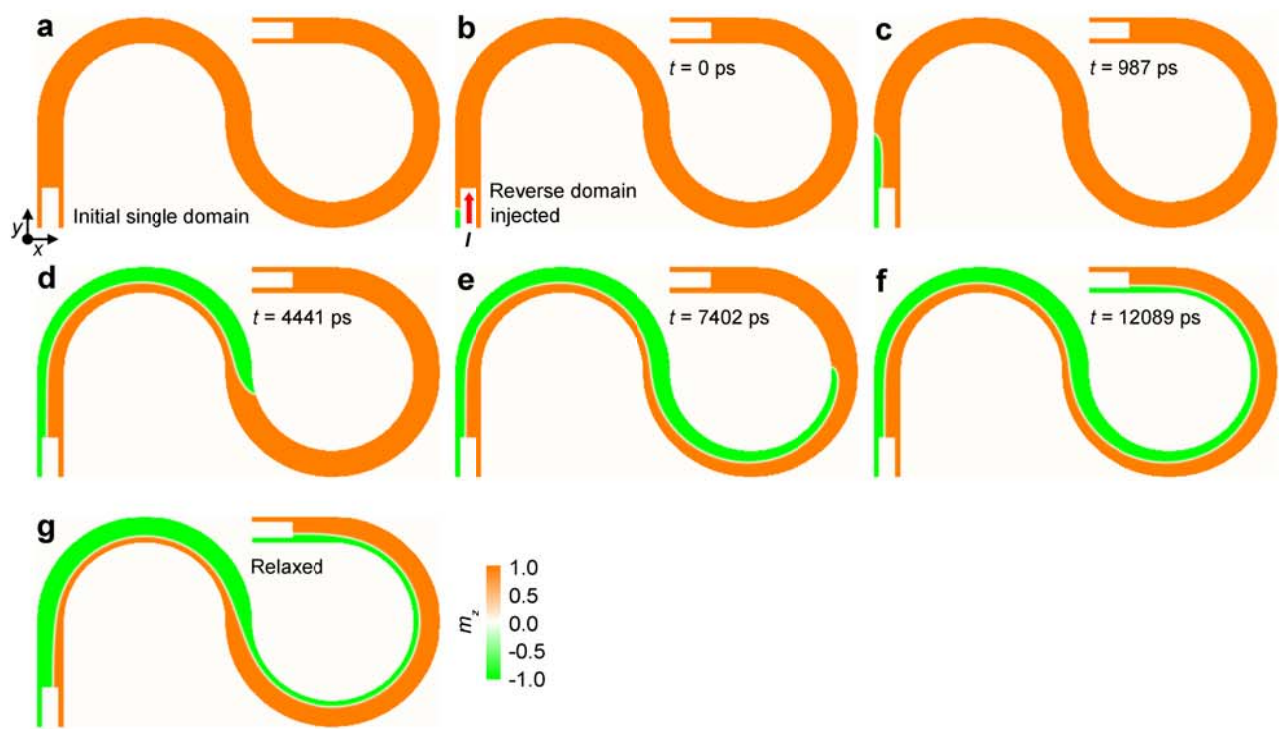

# Paving Spin-Wave Fibers in Magnonic Nanocircuits Using Spin-Orbit Torque

Xiangjun Xing,[1,2] Philip W. T. Pong,[2] J. Åkerman,[3,4] and Yan Zhou[5,*]

[1]School of Physics and Optoelectronic Engineering, Guangdong University of Technology, Guangzhou 510006, China

[2]Department of Electrical and Electronic Engineering, University of Hong Kong, Hong Kong, China

[3]Department of Physics, University of Gothenburg, Fysikgränd 3, 412 96 Gothenburg, Sweden

[4]Material and Nano Physics, School of ICT, KTH Royal Institute of Technology, 164 40 Kista, Sweden

[5]School of Science and Engineering, Chinese University of Hong Kong, Shenzhen, 518172, China

**I. Effect of Rashba Torque and Zhang-Li Torque on SDW Injection**

We examined how the Rashba torque (RsbT) and Zhang-Li torque (STT) affect the injection process of a SDW driven by the spin-Hall torque (SHT). For RsbT, a set of Rashba parameters ($\alpha_R$) ranging from $-1.5\times10^{-21}$ to $1.5\times10^{-21}$ erg·cm were considered with an interval of $0.1\times10^{-21}$ erg·cm. For STT, we consider that $\beta = \alpha$, and we adjust the current direction and the sign of the spin-Hall angle $\Phi_H$ to make STT counteract or strengthen SHT. It is found that, the SDW-injection process would not be significantly affected by RsbT or STT coexisting with SHT under the condition $J_{FM} = J_{HM2}$, if the relative strength of RsbT to SHT is not higher than 0.5. The details are presented in Figures S1 and S2. We also compared the dynamics of SDW and common domain wall; the results are shown in Figures S3.

*E-mail: yanzhou@hku.hk

## II. Effect of Point Impurities on SDW Injection

The nanowires modeled are assumed to be patterned from ultrathin multilayer films, and thus defects can never be totally avoided in real samples. Defects were reported to heavily influence the motion of skyrmions driven by a current [S1]. This fact stimulates us to consider the effect of defects on the SDW injection driven by SHT. The edge irregularities, formed by the staircase along the border of a curved nanowire resulting from finite-difference meshing, are found to contribute negligibly to the SDW injection, as seen from Figures 2 and 3. Here, we consider point impurities randomly distributed inside the nanowire, and find that even for an ultrahigh concentration of impurities, 5%, the SDW injection can still able to be realized satisfactorily. The details are presented in Figure S4. Because of the pinning effect of impurities, the SDWs imprinted into the nanowires containing impurities are not as smooth as that in a perfect wire.

## III. Failure to Inject a Strip Domain with Paired Chiral SDWs

In principle, a strip domain cannot be written into a nanowire by using SHT, because, on one hand, the half-skyrmion attached to the end of the strip domain would distort the SDW (a mechanism proposed recently in Ref. [S2] by Lin) and, on the other hand, one of the paired SDWs of a strip domain will be directly destabilized by SHT (an independent mechanism proposed in the present paper; see Fig. 4). Although we are aware of this fact, we still attempt to inject a strip domain (with paired SDWs) into a wire. However, as expected, no strip domains are injected into the wire, and instead, a SDW is written into the wire. The details are presented in Figure S5.

## IV. Dynamics of a Fractional Skyrmion

A closed strip domain always has two half-skyrmion heads, each of which is tied to an end of the strip domain, whereas a SDW also include a head similar to a quarter-skyrmion. Both the half-skyrmion and quarter-skyrmion carry finite topological charge, and thus are expected to show topological dynamic behavior. Here, the half-skyrmion (quarter-skyrmion) head is bound to the linear part of the entire structure, so that it is impossible to clearly see its topological behavior. We therefore study the current-induced dynamics of a meron (fractional skyrmion) to identify the forces experienced by the SDW head and understand its behavior. The details are presented in Figure S6.

**V. Validity of the Injection Procedure against Geometric Variation**

The injection procedure is valid for Y-shaped nanostructures with 60° and 30° opening angles, as shown in Movies S2 and S3, respectively. Apart from the pad width of $w_N$ = 20 nm, other width values of $w_N$ = 30 and 40 nm were examined, as shown in Movies S4a,b. It was found that, the larger the pad width, the wider the imprinted strip domain (compare Movies S1a, S4a, and S4b); however, following relaxation, the static SDWs are identical. The injection process was tested against a sharp corner around the transition region of the 90° Y-shaped nanostructure (Movie S5) and it can be seen that the SDW head can pass through the sharp corner.

**VI. Validity of the Injection Procedure under Vanishing DMI**

Significantly, this injection procedure can also be applied to samples without DMI, as shown in Movie S6. Comparing Movie S1a ($D$ = 2.0 mJm$^{-2}$) with Movie S6 ($D$ = 0), it can be seen that there is no difference between the injection processes, which indicates that the injection procedure does not rely on the presence of DMI. Nevertheless, SDWs cannot be stabilized in the Y-shaped nanostructure with $D$ = 0; once the writing current is removed, the written SDW deforms rapidly and collapses, resulting in an undesired equilibrium domain pattern, as shown in Figure 5a. This fact highlights the importance of identifying the parameter space where an SDW can exist at static equilibrium.

**VII. Dependence of Static Domain State on Material and Geometric Parameters**

The equilibrium domain patterns in the 90° Y-shaped nanostructure, relaxed from the as-written SDW for various values of $D$ (Fig. 5), clearly indicate that the DMI strength must be in the proper range to stabilize the SDW at static equilibrium (Fig. 5c–e). For subthreshold $D$ values, the as-written SDW will transform into a multidomain texture (Fig. 5a) or will disappear, resulting in a single domain (Fig. 5b); for suprathreshold $D$ values, it will break into the labyrinthine worm-like texture shown in Figure 5f. The stability of SDWs in a Y-shaped nanostructure with respect to DMI strength is different from the situation for a straight magnetic wire, where SDWs are stable even for $D = 0$, as reported in Refs. [S3,S4]. This implies that the stabilization of an SDW in a curved wire requires a sufficiently strong DMI. In the present study, the Y-shaped nanostructure includes a base, two arms, and a transition section, and the transition region between the base and the arm is a segment of a magnetic ring. Therefore, the dependence of SDW stability in a Y-shaped nanostructure on $D$ should be due to the presence of the transition region.

For our Y-shaped nanostructures, the transition region is a 100-nm-wide arc with an outer radius of 400 nm. We thus examined the stability of an SDW in such a 1/4-arc against $M_s$, $K_u$, and $D$; the results are displayed in Figures S7–S9, indicating that an SDW may be stabilized in the 1/4-arc wire at static equilibrium over a broad range of values of $M_s$, $K_u$, and $D$. As expected, the arc has a similar SDW-stability range in $D$ to the Y-shaped nanostructure (compare Fig. 5 and Fig. S9). To clarify the influence of curvature, we checked the SDW stability against $D$ in a set of arcs with the same width (100 nm) but different radii (400, 800, and 1200 nm). We found that the lower threshold value ($D_l^*$) of $D$ required to stabilize an SDW decreases as the radius increases (compare Figs. S9–S11), and for the 1200-nm arc, $D_l^*$ is close to zero (~0.08 mJm$^{-2}$; see Fig. S11). This tendency is consistent with that for straight wires [S3,S4], which are equivalent to arcs with an infinite radius. Surprisingly, the upper threshold values ($D_u^*$) are almost independent of the arc radii (compare Figs. S9–S11).

A detailed knowledge of SDW stability as a function of $K_u$ and $D$ would be useful for device designs, considering that $K_u$ and $D$ are highly sensitive to the interface and layer thickness [S5–S7]. We thus derived the phase diagram for static spin configurations in the 400-nm arc in $K_u$–$D$ space, as shown in Figure S12a. Based on this plot, it is clear that, the larger $K_u$, the wider the $D$ window in which an SDW can be stabilized. For realistic values of $D$, such as $D = 1.0$ mJm$^{-2}$, $K_u$ must be lower than a critical value to give a static SDW; for medium $D$ values, like $D = 2.0$ mJm$^{-2}$, SDWs are stable throughout the entire range of $K_u$; for larger $D$ values, such as $D = 4.0$ mJm$^{-2}$, $K_u$ must be large enough to maintain an SDW. These points can act as a guide to choosing appropriate materials for device applications. Table S1 summarizes the static spin configurations for several special combinations of $K_u$ and $D$, with $M_s$ varying from 80 to 1580 kAm$^{-1}$. These results reveal that an SDW can be stabilized in an arc (a curved wire) over a wide region of the parameter space, which is highly desirable for practical applications of SDW-based magnonic devices [S8].

**VIII. Supplemental Movies**

Eighteen movies are provided separately along with this supplemental PDF file.

MOVIE S1a SDW-injection process achieved by using $I_{dc}$ along S1−S3 to drive an upward seed domain nucleated in pad A of the 90° Y-shaped nanostructure.

MOVIE S1b SDW-injection process achieved by using $I_{dc}$ along S1−S3 to drive a downward seed domain nucleated in pad A of the 90° Y-shaped nanostructure.

MOVIE S1c SDW-injection process achieved by using $I_{dc}$ along S2−S3 to drive a downward seed domain nucleated in pad B of the 90° Y-shaped nanostructure.

MOVIE S1d SDW-injection process achieved by using $I_{dc}$ along S2−S3 to drive an upward seed domain nucleated in pad B of the 90° Y-shaped nanostructure.

MOVIE S1e SDW-injection process achieved by using $I_{dc}$ along S3−S2 to drive an upward seed domain nucleated in pad C of the 90° Y-shaped nanostructure.

MOVIE S1f SDW-injection process achieved by using $I_{dc}$ along S3−S2 to drive a downward seed domain nucleated in pad C of the 90° Y-shaped nanostructure.

MOVIE S1g SDW-injection process achieved by using $I_{dc}$ along S3−S1 to drive a downward seed domain nucleated in pad D of the 90° Y-shaped nanostructure.

MOVIE S1h SDW-injection process achieved by using $I_{dc}$ along S3−S1 to drive an upward seed domain nucleated in pad D of the 90° Y-shaped nanostructure.

MOVIE S2a SDW-injection process achieved by using $I_{dc}$ along S1−S3 to drive an upward seed domain nucleated in pad A of the 60° Y-shaped nanostructure.

MOVIE S2b SDW-injection process achieved by using $I_{dc}$ along S2−S3 to drive a downward seed domain nucleated in pad B of the 60° Y-shaped nanostructure.

MOVIE S3a SDW-injection process achieved by using $I_{dc}$ along S1−S3 to drive an upward seed domain nucleated in pad A of the 30° Y-shaped nanostructure.

MOVIE S3b SDW-injection process achieved by using $I_{dc}$ along S2−S3 to drive a downward seed domain nucleated in pad B of the 30° Y-shaped nanostructure.

MOVIE S4a SDW-injection process achieved in the 90° Y-shaped nanostructure by using a 30-nm-wide injection pad.

MOVIE S4b SDW-injection process achieved in the 90° Y-shaped nanostructure by using a 40-nm-wide injection pad.

MOVIE S5 SDW-injection process realized in the 90° Y-shaped nanostructure with a sharp corner around the transition section.

MOVIE S6 SDW-injection process realized in the 90° Y-shaped nanostructure with vanishing DMI (*i.e.*, $D = 0$).

MOVIE S7a SDW-injection process realized in the 90° Y-shaped nanostructure at room temperature.

MOVIE S7b SDW-injection process realized in the 90° Y-shaped nanostructure with impurities (concentration $c = 1\%$) at room temperature.

**IX. Supplemental Figures and Table**

Twelve figures (Fig. S1–S12) and 1 table (Table S1) are included in this PDF file.

FIG. S1

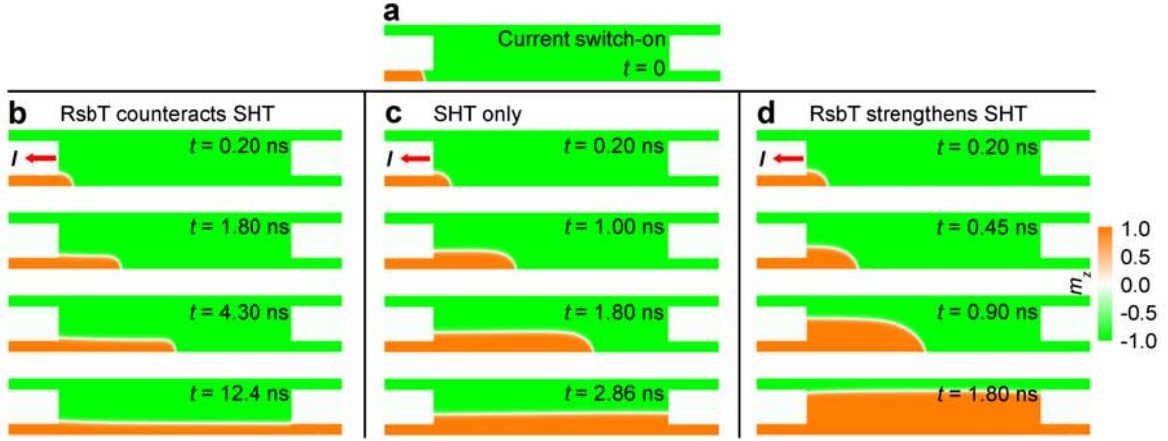

FIG. S1 Effect of Rashba torque (RsbT) coexisting with the spin-Hall torque (SHT) on SDW injection. Here, $D = -2$ mJm$^{-2}$, $\Phi_H = +0.13$, $P = 0.4$, and we assumed that $J_{FM} = J_{HM2} = 3\times10^{12}$ Am$^{-2}$. (a) Seed reverse domain nucleated at the lower-left pad. Current is turned on at the time $t = 0$ ns when the seed domain is just formed. Snapshots of the SDW at various times after the current action for (b) $\alpha_R = +0.5\times10^{-21}$ erg·cm, (c) $\alpha_R = 0$, and (d) $\alpha_R = -0.5\times10^{-21}$ erg·cm.

Note that for $\alpha_R = \pm0.5\times10^{-21}$ erg·cm, the relative magnitude of RsbT to SHT is $|\tau_R/\tau_H| = 0.5$. The Rashba torque induces transverse shift of the SDW ridge and suppresses (strengthens) the drift motion of the SDW head along the edge for a positive (negative) Rashba parameter $\alpha_R$.

FIG. S2

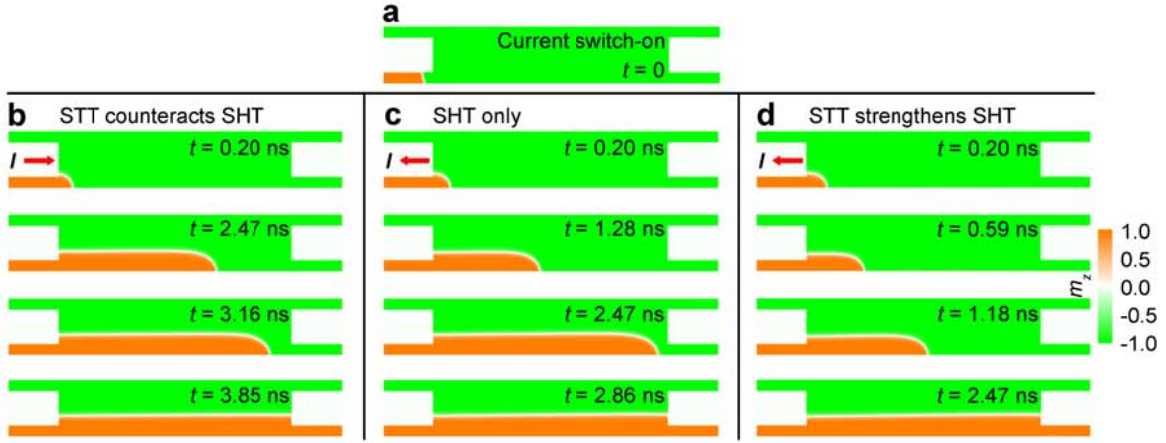

FIG. S2 Effect of Zhang-Li spin-transfer torque (STT) coexisting with the spin-Hall torque (SHT) on SDW injection. Here, $D = -2$ mJm$^{-2}$, $P = 0.4$, and we assumed $\beta = \alpha$. (a) Seed reverse domain nucleated at the lower-left pad. Current is turned on at the time $t = 0$ ns when the seed domain is just formed. Snapshots of the SDW at various times after the current action for (b) $\Phi_H = -0.13$ and $J_{FM} = J_{HM2} = 3\times10^{12}$ Am$^{-2}$ along +$x$, (c) $\Phi_H = +0.13$, $J_{FM} = 0$, and $J_{HM2} = 3\times10^{12}$ Am$^{-2}$ along -$x$, and (d) $\Phi_H = +0.13$ and $J_{FM} = J_{HM2} = 3\times10^{12}$ Am$^{-2}$ along -$x$.

Note that STT can suppress or enhance SHT, depending on the direction of applied current and the spin-Hall angle.

FIG. S3

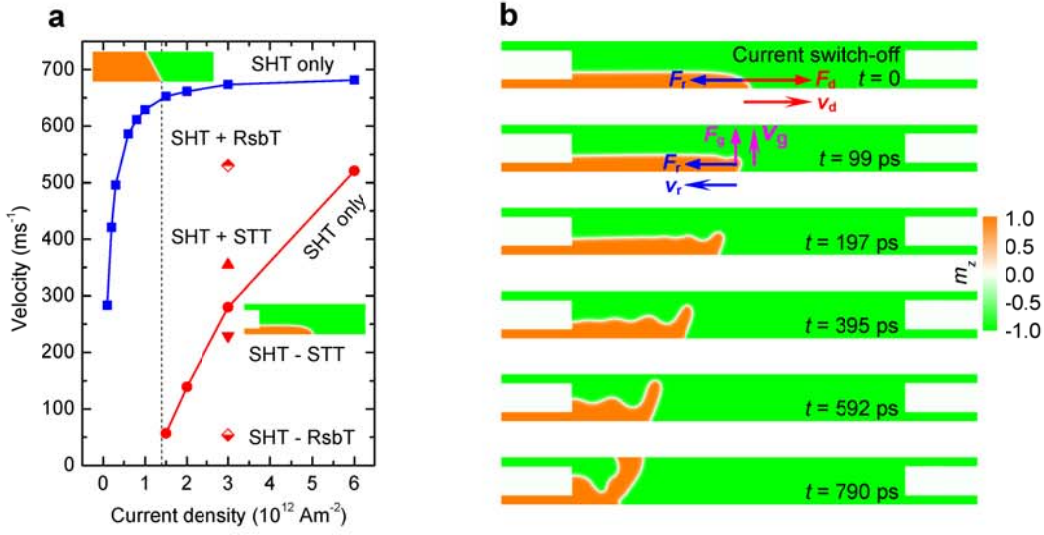

FIG. S3 (a) Domain-wall velocity and SDW-head velocity versus current density. The blue squares represent the velocity values of the domain wall subjected to SHT only. The red circles denote the velocity values of the SDW subjected to SHT only. The dashed vertical line labels the critical current density, $J_c \approx 1.35 \times 10^{12}$ Am$^{-2}$, above which the SDW head in the nanowire can be moved by SHT. The triangles give the velocity values of the SDW under coexisting SHT and STT. SHT+STT (SHT-STT) indicates that STT enhances (suppresses) the SDW motion under SHT. The diamonds stand for the velocity values of the SDW under coexisting SHT and RsbT. SHT+RsbT (SHT-RsbT) indicates that RsbT enhances (suppresses) the SDW motion under SHT. Inserted into the graph are the domain wall (top) and SDW (bottom). $J_{FM} = J_{HM2}$ was assumed. (b) Relaxation dynamics of a SDW. The current ($J_{HM2} = 3 \times 10^{12}$ Am$^{-2}$) suddenly stops at the time $t = 0$ ps when the SDW head is on the half way to the right end of the nanowire. All data in (a) and (b) were simulated with $D = -2$ mJm$^{-2}$.

In Figure S3a, below $J_c$, the SHT-related force cannot overcome the elastic restoring force, and the SDW head in the nanowire does not move. At $J_c$, the elastic restoring force is just balanced by the SHT-related force, and the SDW head will be kept stationary. Above $J_c$, the SDW head in the nanowire can be displaced by SHT. Note that $J_c$ does not apply to the cases where RsbT or STT accompanies SHT. As seen from Figure S3a, the critical current density for the SDW is much higher than that for the common domain wall, which should be attributed to the higher elastic energy of the former, as revealed in Figure S3b. We did not see auto-oscillation in SDWs in the covered current-density range up to $6 \times 10^{12}$ Am$^{-2}$ [S9].

In Figure S3b, before the current stops, the current-induced force $F_d$ on the SDW head is balanced by the restoring force $F_r$, and the SDW head moves steadily at a constant speed $v_d$. Once the current is removed, $F_d$ disappears, and the SDW head contracts rapidly, at a negative speed $v_r$, under the elastic restoring force originating from the high elastic energy of the SDW. The speed $v_r$ causes a transverse Magnus force on the SDW head because of the finite topological charge of the latter. In turn, the Magnus force drags the SDW head upward, and eventually, the ordered SDW structure is destroyed. This simulated dynamic process is consistent with Thiele's theory [S10].

FIG. S4

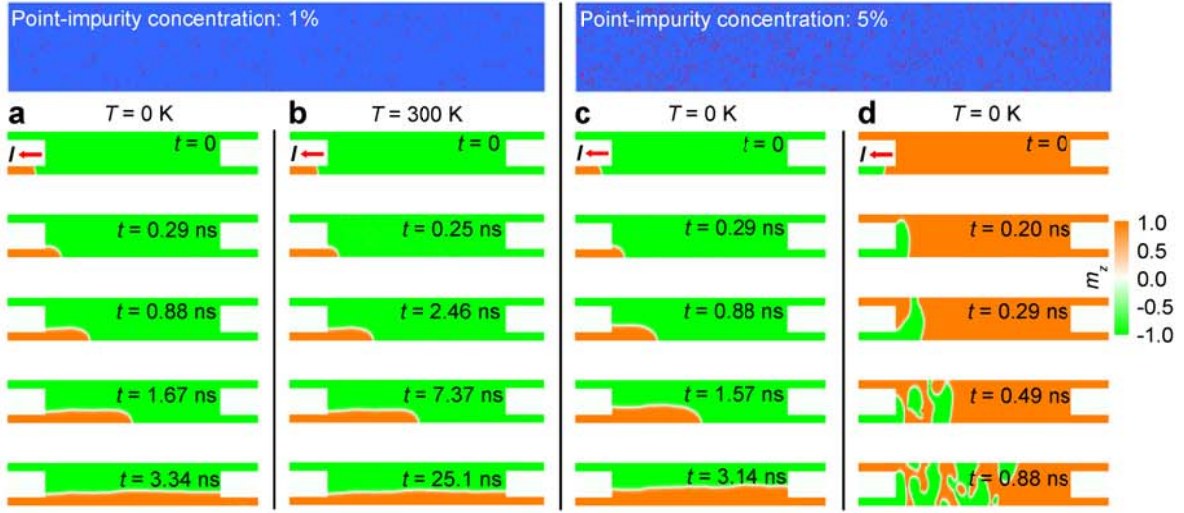

FIG. S4 Effect of impurities on the SDW injection driven by SHT. The top-left and top-right panels display the impurity distributions for the impurity concentration $c = 1\%$ and $5\%$, respectively. $D = -2$ mJm$^{-2}$ for the non-impurity sites, whereas $D = 2$ mJm$^{-2}$ for the impurity sites. $J_{HM2} = 3\times10^{12}$ Am$^{-2}$. SDW injection process for the $c = 1\%$ nanowire at the temperature (a) $T = 0$ K and (b) $T = 300$ K. SDW injection process for the $c = 5\%$ nanowire at $T = 0$ K initiated from (c) a right seed domain and (d) a wrong seed domain.

Comparing Figure S4, Figure 2, Figure 3, and Movie S7, it is clear that a SDW can be successfully written into a nanowire if a correctly set seed domain is used, irrespective of the geometry of the nanowire (straight or curved), the defect concentration in the wire (with or without impurities), and the operating temperature (0 or 300 K).

FIG. S5

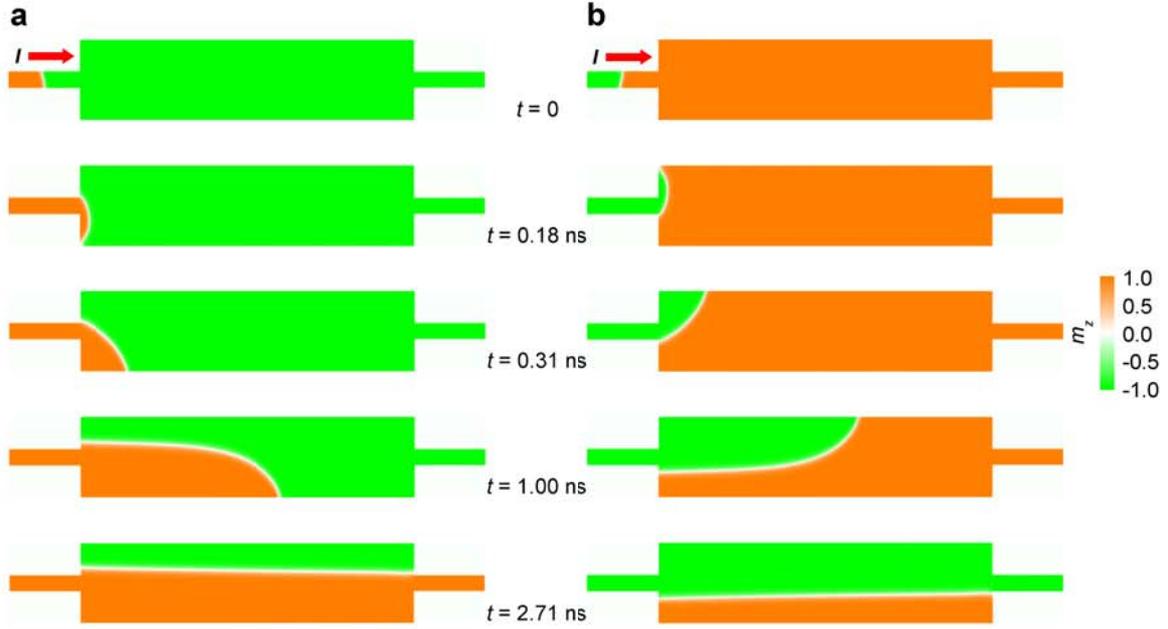

FIG. S5 Failure to inject a strip domain with paired chiral SDWs. Here, $D = 2$ mJm$^{-2}$ and $J_{HM2} = 3 \times 10^{12}$ Am$^{-2}$. The nucleation pads are placed along the central axis of the sample to maintain the mirror symmetry. The seed domains in (a) and (b) are upward and downward, respectively. From $t = 0$ on, the current is applied.

In each case, a SDW instead of a strip domain, with paired SDWs, is imprinted into the wire. Obviously, only the SDW, in which $\mathbf{m} \parallel \boldsymbol{\sigma}$, is permitted to reside in the nanowire, and the other SDW, in which $\mathbf{m}$ is antiparallel to $\boldsymbol{\sigma}$, is always expelled from the nanowire. Consequently, the desired strip domain, with paired SDWs, can never form, regardless of the orientation of the seed domain.

FIG. S6

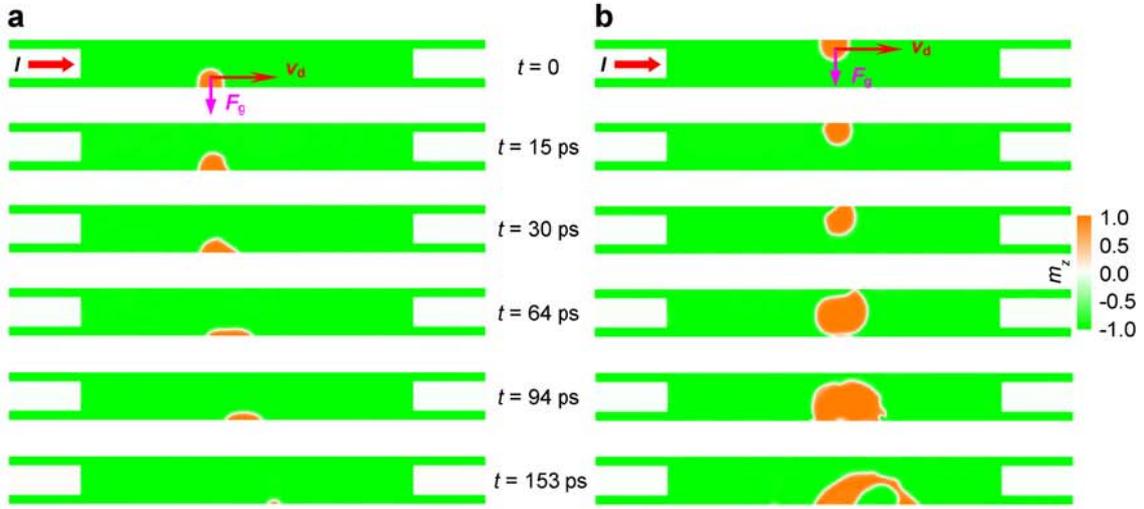

FIG. S6 Dynamics of a meron — a magnetic object with fractional topological charge. Here, $D = 3.5$ mJm$^{-2}$ and $J_{HM2} = 3 \times 10^{12}$ Am$^{-2}$. Initially, the meron sits at (a) the bottom edge and (b) the top edge of the nanowire. Under the current, the meron moves along the edge and thus acquires a longitudinal speed $v_d$. The longitudinal motion then causes a Magnus force $F_g$ to act on the meron owing to the meron's nonzero topological charge. The Magnus force expels the meron from the nanowire in (a) and drags the meron into the nanowire in (b).

These current-induced forces experienced here by the standalone meron should also occur to the half-skyrmion attached to the strip domain (Fig. 4a) and occur to the quarter-skyrmion head bound to the SDW (Fig. 4b). Therefore, the SDW head in the top panel of Fig. 4b feels an outward force, which stabilizes the SDW; whereas the SDW head in the bottom panel of Fig. 4b undergoes an inward force, which distorts the SDW according to Lin's theory [S2].

FIG. S7

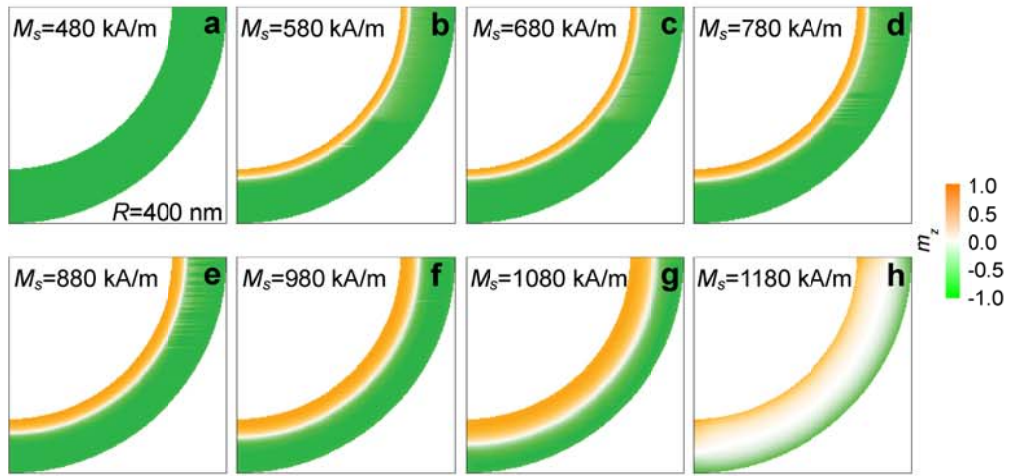

FIG. S7 Static domain pattern in a curved wire (1/4-arc) as a function of $M_s$. Other parameters are fixed at $A$=15 pJm$^{-1}$, $K_u$=0.8 MJm$^{-3}$, and $D$=1.0 mJm$^{-2}$. The width and outer radius of the wire are 100 and 400 nm, respectively.

FIG. S8

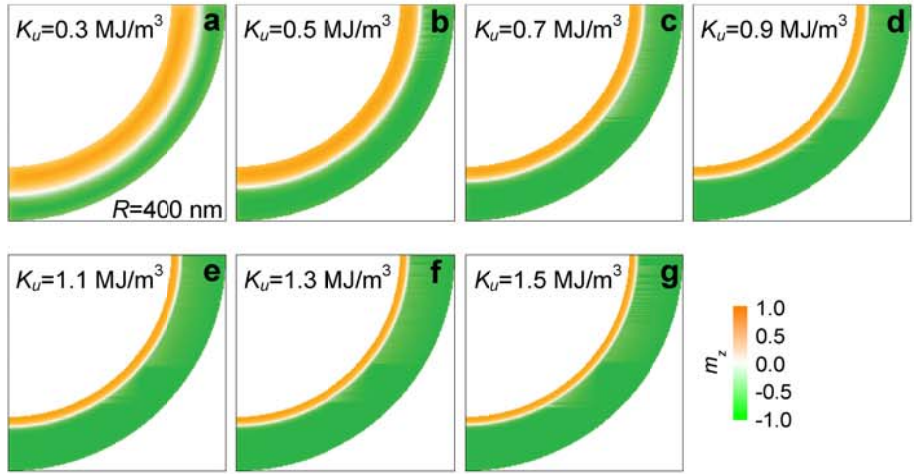

FIG. S8 Static domain pattern in a curved wire (1/4-arc) as a function of $K_u$. Other parameters are fixed at $A$=15 pJm$^{-1}$, $M_s$=580 kAm$^{-1}$, and $D$=2.0 mJm$^{-2}$. The width and outer radius of the wire are 100 and 400 nm, respectively.

FIG. S9

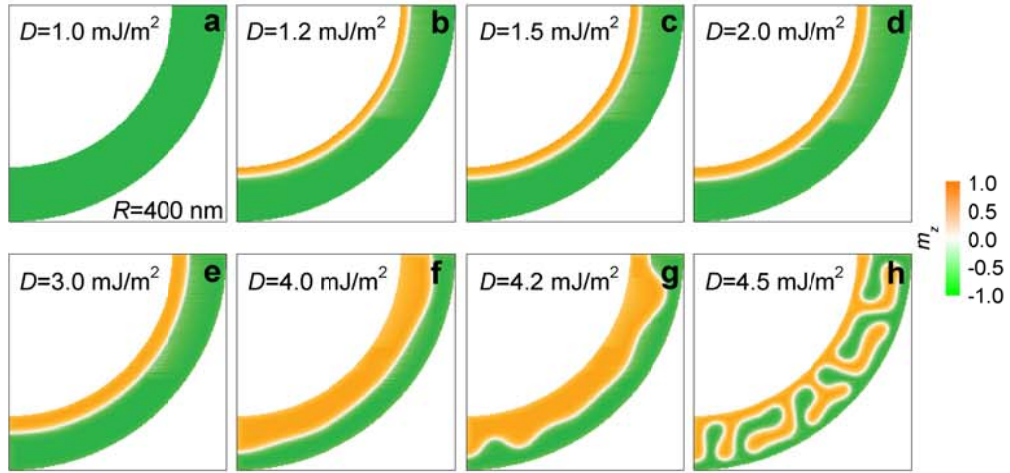

FIG. S9 Static domain pattern in a curved wire (1/4-arc) as a function of $D$. Other parameters are fixed at $A$=15 pJm$^{-1}$, $M_s$=580 kAm$^{-1}$, and $K_u$=0.8 MJm$^{-3}$. The width and outer radius of the wire are 100 and 400 nm, respectively.

FIG. S10

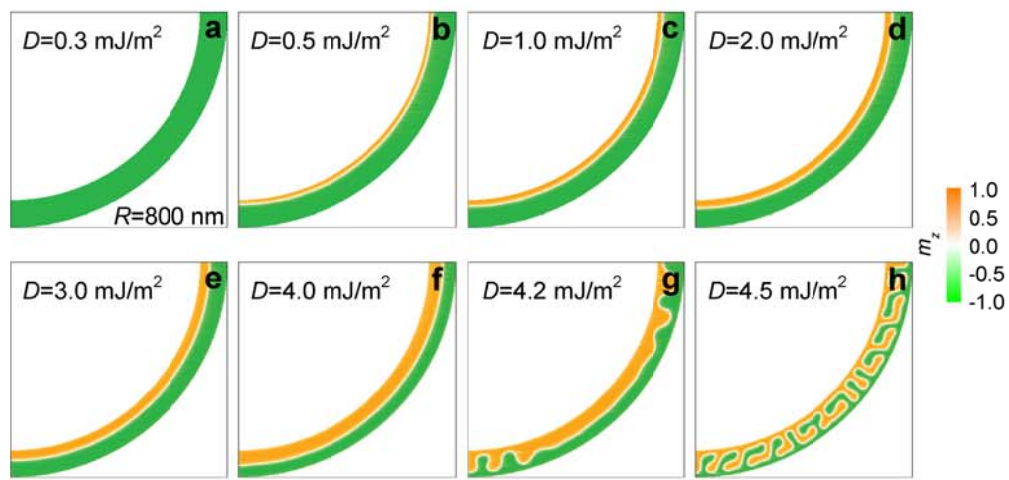

FIG. S10 Static domain pattern as a function of $D$ in a curved wire (1/4-arc) with an increased outer radius of 800 nm. Other parameters are fixed at $A$=15 pJm$^{-1}$, $M_s$=580 kAm$^{-1}$, and $K_u$=0.8 MJm$^{-3}$. The width of the wire is 100 nm.

FIG. S11

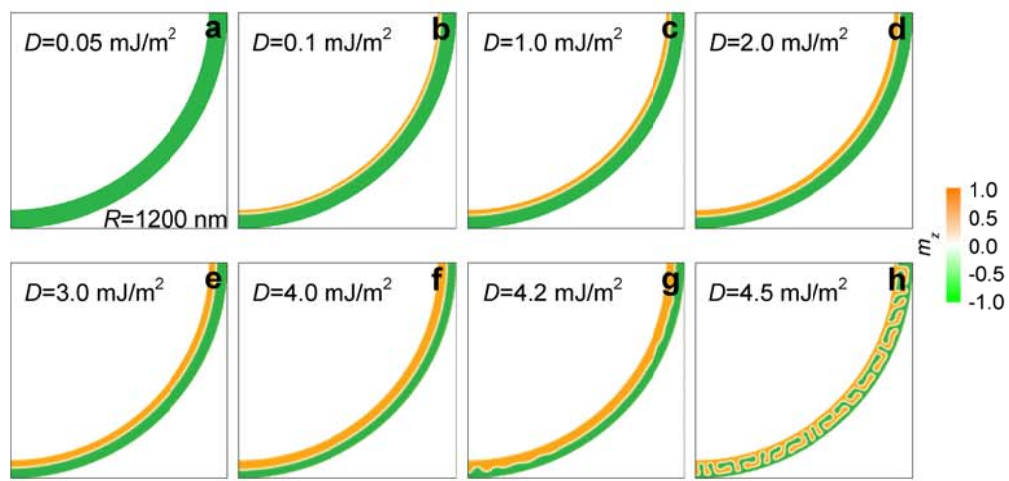

FIG. S11 Static domain pattern as a function of $D$ in a curved wire (1/4-arc) with an increased outer radius of 1200 nm. Other parameters are fixed at $A$=15 pJm$^{-1}$, $M_s$=580 kAm$^{-1}$, and $K_u$=0.8 MJm$^{-3}$. The width of the wire is 100 nm.

FIG. S12

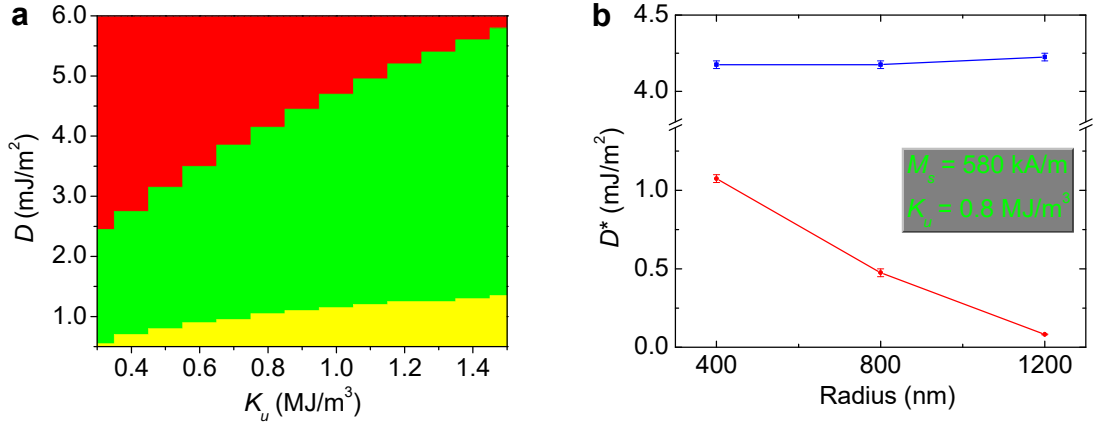

FIG. S12 (a) Phase diagram for static magnetization states in a curved wire (1/4-arc) in the $K_u$–$D$ space. $R$=400 nm, $w$=100 nm, $A$=15 pJm$^{-1}$, and $M_s$=580 kAm$^{-1}$. The bottom, middle, and top regions correspond to single-domain, SDW, and multi-domain configurations, respectively, and the characteristic domain patterns are those in Fig. S9a,d,h. (b) Critical DMI strengths ($D^*$) versus the outer radius of a curved wire (1/4-arc). $w$=100 nm, $A$=15 pJm$^{-1}$, $M_s$=580 kAm$^{-1}$, and $K_u$=0.8 MJm$^{-3}$. Square and circular dots denote the upper ($D_u^*$) and lower ($D_l^*$) critical values, respectively. Error bars are marked along with data points. Apparently, the smaller the curved wire is, the higher the $D_l^*$. A SDW can be stabilized at static equilibrium when $D_l^*<D<D_u^*$.

Table S1

| $M_s$ \ $(K_u, D)$ | 80 | 180 | 280 | 380 | 480 | 580 | 680 | 780 | 880 | 980 | 1080 | 1180 | 1280 | 1380 | 1480 | 1580 |
|---|---|---|---|---|---|---|---|---|---|---|---|---|---|---|---|---|
| (0.8, 1.0) | 0 | 0 | 0 | 0 | 0 | 1 | 1 | 1 | 1 | 1 | 1 | 2 | 2 | 2 | 2 | 2 |
| (0.8, 3.5) | 1 | 1 | 1 | 1 | 1 | 1 | 1 | 1 | 2 | 2 | 2 | 2 | 2 | 2 | 2 | 2 |
| (1.4, 1.0) | 0 | 0 | 0 | 0 | 0 | 0 | 0 | 0 | 1 | 1 | 1 | 1 | 1 | 1 | 1 | 2 |
| (1.4, 3.5) | 1 | 1 | 1 | 1 | 1 | 1 | 1 | 1 | 1 | 1 | 1 | 1 | 1 | 2 | 2 | 2 |

Table S1 Static magnetization states in a curved wire (1/4-arc) as a function of $M_s$, $K_u$, and $D$. The units of $M_s$, $K_u$, and $D$ are kAm$^{-1}$, MJm$^{-3}$, and mJm$^{-2}$, respectively. The numbers 0, 1, and 2 represent the single-domain, SDW, and multi-domain states, respectively. $A$=15 pJm$^{-1}$, $R$=400 nm, and $w$=100 nm.